\documentclass[pra,10pt,twocolumn,reprint,tikz,floatfix,showpacs]{revtex4-2}
\usepackage[utf8]{inputenc}
\usepackage{amsthm}
\usepackage{comment}
\usepackage[T1]{fontenc}    
\usepackage[british]{babel}  
\usepackage[sc,osf]{mathpazo}
\usepackage[scaled=0.86]{berasans}  
\usepackage[colorlinks=true, citecolor=blue, urlcolor=blue]{hyperref}  
\usepackage{float}
\usepackage{graphicx} 
\usepackage{subfig}
\usepackage{bitset}
\usepackage[babel]{microtype}  
\usepackage{tikz}
\usetikzlibrary{arrows.meta, positioning,shapes,calc}
\usepackage{amsmath,amssymb,amsthm,bm,amsfonts,mathrsfs,bbm} 
\usepackage{xspace}  
\usepackage{pgfplots}
\usepackage{xcolor,colortbl}
\def\ba{\begin{equation}}
	\def\ea{\end{equation}}
\def\bea{\begin{eqnarray}}
	\def\eea{\end{eqnarray}}
\def\ben{\begin{equation*}}
	\def\een{\end{equation*}}
\def\bean{\begin{eqnarray*}}
	\def\eean{\end{eqnarray*}}
\def\bma{\begin{mathletters}}
	\def\ema{\end{mathletters}}
\def\bi{\begin{itemize}}
	\def\ei{\end{itemize}}

\newcommand{\be}{\begin{equation}}
	\newcommand{\ee}{\end{equation}}

\newcommand{\kommentar}[1]{}

\newcommand{\forget}[1]{}

\newtheorem{theorem}{Theorem}

\begin{document}
	
	\title{Measurement Incompatibility Based In-equivalence Between Bell and Network Nonlocality}
	\author{Kaushiki Mukherjee}
	\email{kaushiki.wbes@gmail.com}
	\affiliation{Department of Mathematics, Government Girls' General Degree College, Ekbalpore, Kolkata-700023, India.}
	\author{Biswajit Paul}
	\email{biswajitbbkm01082019@gmail.com}
	\affiliation{Department of Mathematics, Balagarh Bijoy Krishna Mahavidyalaya, Hooghly, West Bengal, India.}
	
\begin{abstract}
It is a well-known fact that measurement incompatibility is a necessary resource to generate nonlocal correlations in usual Bell scenario that typically involves single quantum source. We can provide with some contrasting findings if we consider connected structure of multiple quantum sources. Precisely, we demonstrate that non $n$-locality can be detected in standard quantum network even when only a single party performs incompatible measurements. More interestingly, for any finite $n$$\geq$$3,$ non $n$-local correlations can be generated in any standard linear $n$-local network when all the parties perform compatible measurements. Such an observation is topology specific as one of the parties must perform incompatible measurement to exhibit non $n$-locality in any non-linear network endowed with star topology. However, we observe that in any non-standard network(all sources independent and nonlocal), to generate genuine non $n$-local correlations, all the parties must perform incompatible measurements. Such a finding is intuitive as more resource is required to generate stronger form of quantum non-classicality. We also demonstrate that merely providing resource of measurement incompatibility to all the parties is not sufficient for non $n$-locality detection in any quantum network.
\end{abstract}
	
	\maketitle
	
	
Any quantum network, being a connected structure of multiple parties and sources, is commonly expected to have enough potential to generate new notion of nonlocal quantum correlations that are inexplicable in set-ups associated with standard Bell scenarios \cite{Bci,Bran,Fri,Ros,Cha,Ren,Reno, Kauu,Kas,Muk,Taav}. However, recent studies cast doubt over this belief based on findings that the observed nonlocality in a standard $n$-local network, can still be traced back to Bell-CHSH nonlocality \cite{Cla} existing in some individual pairs of parties(forming nodes in the network) \cite{Gis,And,Tavak,Poz}. For instance, violation of BRGP inequality(\cite{Bran}) can be interpreted in terms of Bell-CHSH inequality violation observed in individual pairs of nodes in standard $n$-local network \cite{Tavak}. Again, in any such network, if one out of $n$ pairs of nodes exhibit Bell-CHSH nonlocality, $n+1$-partite non $n$-local correlations can be generated, regardless of the nature of bipartite correlations in between nodes comprising each of remaining $n-1$ pairs\cite{Poz}. 
\begin{figure}
	\centering
	\begin{tikzpicture}[
		node distance=1.2cm and 1cm,
		every node/.style={font=\boldmath\bfseries\small},
		meas/.style={draw, circle, minimum size=0.55cm,inner sep=0pt,       
			align=center, 
			text height=1.2ex, 
			text depth=0ex},
		source/.style={draw, rectangle, minimum width=0.5cm, minimum height=0.3cm},
		arr/.style={->, thick},
		dots/.style={gray, thick}
		]
		
		\node[meas] (A1) at (0,0) {$\,A_{1\,\,}$};
		\node[meas] (A2) [right=of A1] {$\,A_{2\,\,}$};
		\node[meas] (A3) [right=of A2] {$\,A_{3\,\,}$};
		\node at ($(A3)+(0.6,0)$) {\dots};
		\node[meas] (An) at ($(A3)+(1.2,0)$) {$\,A_{n\,\,}$};
		\node[meas] (Anp1) [right=of An] {$\mathbf{\small{A_{n+1}}}$};
		
		\node[source] (S1) [below=0.5cm of $(A1)!0.4!(A2)$] {$\mathcal{S}_1$};
		\node[source] (S2) [below=0.5cm of $(A2)!0.4!(A3)$] {$\mathcal{S}_2$};
		\node[source] (Sn) [below=0.5cm of $(An)!0.4!(Anp1)$] {$\mathcal{S}_n$};
		
		\draw[arr] (S1) -- (A1);
		\draw[arr] (S1) -- (A2);
		
		\draw[arr] (S2) -- (A2);
		\draw[arr] (S2) -- (A3);
		
		\draw[arr] (Sn) -- (An);
		\draw[arr] (Sn) -- (Anp1);
		
		\draw[arr] (A1) -- ++(0,-0.9) node[below] {$a_1$};
		\draw[arr] (A2) -- ++(0,-0.9) node[below] {$\vec{a}_2$};
		\draw[arr] (A3) -- ++(0,-0.9) node[below] {$\vec{a}_3$};
		\draw[arr] (An) -- ++(0,-0.9) node[below] {$\vec{a}_n$};
		\draw[arr] (Anp1) -- ++(0,-0.9) node[below] {$\mathbf{\small{a_{n+1}}}$};
		
		\node at ($(A1)+(0,1)$) {$\{\mathbf{\small{x_{1,j}}}\}$};
		\draw[arr] ($(A1)+(0,0.8)$) -- (A1);
		
		\node at ($(Anp1)+(0,0.9)$) {$\{\,\mathbf{\small{x_{n+1,j}}}\}$};
		\draw[arr] ($(Anp1)+(0,0.8)$) -- (Anp1);\
	\end{tikzpicture}
	\caption{Schematic Diagram of linear $n$-local network}
	\label{nlocal1}
\end{figure}
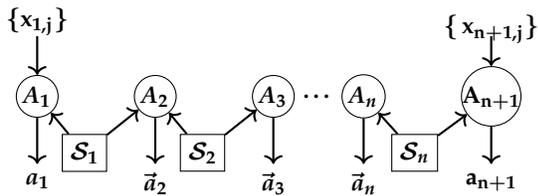
	All these results point out that network nonlocality in any standard $n$-local network cannot be considered as a truly network phenomenon. Recently, to explore novel form of network non-classicality, notion of full network nonlocality has been introduced in \cite{Poz}. Such a notion forbids interpretation of full network nonlocality in terms of standard bipartite Bell nonlocality in any pair of nodes. In any such network topology, none of the sources is of local variable nature\cite{Poz}. This is a more stronger restriction over the sources compared to that imposed in any standard $n$-local network topology. From the perspective of this stronger restriction and also for ease of discussion, we can safely refer to a $n$-local network involving no local source as \textit{non standard $n$-local network} in our further discussions. \\
	Recent findings in study of quantum networks point out the requirement of designing non standard $n$-local networks\cite{Taav,Poz}, hence, discouraging use of standard $n$-local networks for exploiting intrinsic network features in context of witnessing quantum nonlocality. Present work explore the possibility(if any) to identify some feature of standard $n$-local networks that is intrinsically dependent on the connected structure of the nodes and the sources. Precisely speaking, present work tries to establish in-equivalence in between the task of detecting non $n$-locality in network and that of detecting standard Bell nonlocality individually in any pair of nodes. Interestingly, it turns out that such in-equivalence emerges in perspective of quantum measurement incompatibility. \\
	Measurement incompatibility, the impossibility of jointly measuring certain observables,  is a fundamental signature of non-classicality in quantum theory. Two or more measurements are said to be incompatible if there exists no single parent measurement (a joint POVM) from which their outcome statistics can be recovered via classical post-processing \cite{Hein,Guh}. In the context of standard Bell nonlocality, incompatible measurements are necessary for demonstrating violation of any Bell inequality \cite{Fin,Wol}. However,for general bipartite scenarios, it was proven that there are sets of measurements which are incompatible, but cannot lead to Bell nonlocality \cite{Qui, Hir, Ben}.\\
	First let us assume that anyone of $\rho_1,\rho_2$ is distributed among two distant parties $A_1$ and $A_{2}.$ To detect Bell-CHSH nonlocality, using $\rho_1$ or $\rho_2$ individually, both the parties must parties must choose from set of 2 incompatible measurements\cite{Fin,Wol}. Again, let both $\rho_1,\rho_2$ be shared among $A_1,$ $A_{2}$ simultaneously such that $A_1$ and $A_{2}$ both receive two particles(one from each $\rho_1,\rho_2$). By using some entanglement concentration protocol over $\rho_1\otimes \rho_2$, $A_1,A_2$ may share a maximally entangled state\cite{Benn} $\rho_{ent}$(say). To detect Bell nonlocality from $\rho_{ent},$ both $A_1$ and $A_{2}$ need to perform incompatible measurements \cite{Fin,Wol}.\\  Now, let us consider a bilocal network set-up(Fig.\ref{nlocal1} for $n$$=$$2$).
Let source $\mathcal{S}_i$($ i$$=$$1,2$) distribute two-qubit entangled state $\rho_i$ between parties $A_i,A_{i+1}$. Here, non bilocality of network correlations are detected via violation of BRGP inequality\cite{Bran}. As already discussed, such violation can be interpreted in terms of Bell-CHSH violation individually by $\rho_1$ and $\rho_2$. Such results provide intuitions that each party must perform incompatible measurements to generate non bilocal correlations. However, we will report some counter-intuitive findings. Let us first provide an example.\\
Let $\mathcal{S}_i$ distribute Werner state $v_i(|\psi^-\rangle\langle \psi^-|)+(1-v_i)\frac{\mathbb{I}_{2\times 2}}{4}$ with $v_1$$=$$0.87$ and $v_2$$=$$0.97$ respectively. Let $A_2$ perform single Bell basis measurement; $A_1$ choose from set $\{x_{1,0},x_{1,1}\}$ of two incompatible measurements(Appendix.\ref{appna}) $x_{1,i}$$=$$0.664(\sigma_3+(-1)^i\sigma_1)$ and $A_3$ perform from set $\{x_{3,0},x_{3,1}\}$ of two compatible measurements: $x_{3,i}$$=$$0.494(\sigma_3+(-1)^i\sigma_1).$
	Resulting measurement correlations are used to test BRGP inequality \cite{Bran}. This inequality is given by following $n$-local inequality\cite{Bran,Kauu} for $n$$=$$2$:
	\begin{eqnarray}\label{ineqb}
		&& \sqrt{|I_n|}+\sqrt{|J_n|}\leq  1,\,  \textmd{where}\\
		&& I_n=\frac{1}{4}\sum_{i,j=0}^1\langle \small{D}_{1,x_{1,i}}D_2^0D_3^0...\small{D}_{n}^0D_{n+1,x_{n+1,j}}\rangle\nonumber\\
		&&   J_n=\frac{1}{4} \sum_{i,j=0}^1 (-1)^{i+j}\langle \small{ \small{D}_{1,x_{1,i}}D_2^1...\small{D}_{n}^1D_{n+1,x_{n+1,j}}}\rangle\,\,\textmd{with} \nonumber\\
		&&   \langle \small{D}_{1,x_{1,i}}D_2^k...\small{D}_{n}^kD_{n+1,x_{n+1,j}}\rangle = \sum_{\mathcal{D}_1} (-1)^{a_1+a_{n+1}+\sum_{j=2}^{n}a_{j (k+1)}}Q,\nonumber\\
		&& \textmd{\small{where}}\,Q=\small{p (a_1,
			\bar{a}_2,...,\bar{a}_{n},
			a_{n+1}|x_{1,k},x_{n+1,k})},\,\, k=0,1\nonumber\\
		&& \mathcal{D}_1=\{a_1,a_{21},a_{22},...,
		a_{n1},a_{n2},a_{n+1}\}\nonumber\\
		&&\bar{a}_i=(a_{i1},a_{i2})
	\end{eqnarray}
	
	Here L.H.S. of BRGP inequality gives $1.013.$ Non bilocality is thus detected in the network when only $A_1$ performs incompatible measurements.
	So, from operational view point, network nonlocality detection task is not equivalent to Bell nonlocality detection task even though both the tasks involve the same entangled sources. This is due to lesser requirement of resource(in terms of measurement incompatibility) for detecting non bilocality. Such an in-equivalence in nonlocality detection may be attributable to the network structure that allows simultaneous distribution of particles from two entangled sources suitably among $3$ parties(nodes) in contrast to using $\rho_1,\rho_2$ individually or first generating better entangled state $\rho_{ent}$ from $\rho_1\otimes \rho_2$(via entanglement concentration protocol) and then using $\rho_{ent}$ in Bell scenario. Also, as per the network set-up, the central party receives two qubits thereby getting chance to measure in maximally entangled basis.\\
	\par Above example can be generalized to scenario involving $n$ independent sources $\mathcal{S}_1,...,\mathcal{S}_n$. Let us formalize our findings encompassing above example.\\
	\textbf{Network Scenario:} Consider any linear $n$-local network(Fig.\ref{nlocal1}) involving $n$ independent sources $\mathcal{S}_1,...,\mathcal{S}_n$.  $\forall i,\mathcal{S}_i$ distributes particles among two parties(nodes) $A_{i},A_{i+1}.$ Each of $A_2,...,A_n$ has single input whereas each of the extreme parties $A_1$ and $A_{n+1}$ has two inputs. $n+1$-partite measurement correlations are $n$-local if those can be factorized in terms of the local hidden variables $\lambda_i$ characterizing sources $\mathcal{S}_i(i$$=$$1,...,n)$:
	\begin{eqnarray}\label{nlocal}
		P(a_1,\bar{a}_2,..,\bar{a}_n,a_{n+1}|x_1,x_{n+1})=\sum_{\lambda_1\in\Lambda_1}..\sum_{\lambda_n\in\Lambda_n} \mu(\lambda_1,...,\lambda_n)\mathcal{P}_1&&\nonumber\\ \textmd{\small{with}}\,\mathcal{P}_1=P(a_1|x_1,\lambda_1)\Pi_{i=2}^{n}P(\bar{a}_i|\lambda_{i-1},\lambda_i)P(a_{n+1}|x_{n+1},\lambda_n)&&\nonumber\\
		&&\\
		\textmd{\small{and n-local constraint:}}\,\,\mu(\lambda_1,...,\lambda_n)=\Pi_{i=1}^n \mu_i(\lambda_i)&&\nonumber\\
		&&
	\end{eqnarray}
	
	Correlations inexplicable in above form are non $n$-local. Violation of existing $n$-local inequality ensures non $n$-locality of the network correlations.\\ 
	Now, let each of the sources distribute arbitrary two-qubit state. In such a network scenario, we observe an interesting result:
	\begin{theorem}\label{thinc1}
		In any standard linear $n$-local network, non $n$-locality can be detected even if both the extreme parties do not perform incompatible measurements.
	\end{theorem}
	\textit{Proof:} See Appendix.\ref{appnb} for the proof.\\
	Theorem.\ref{thinc1} ensures existence of quantum sources, incompatible measurement settings for one extreme party and compatible measurements for the other for which non $n$-locality can be detected in any linear $n$-local network(Appendix.\ref{appna}). In course of proving the theorem, we characterize quantum sources along with the measurement contexts involving noisy Von-Neumann equatorial settings for which above result holds. \\
	Theorem.\ref{thinc1} is in contrast to existing result in Bell scenario where more resource is required to display nonlocality as incompatibility of measurements is necessary for all the parties involved therein. However, lesser requirement of resource to detect non $n$-locality is not applicable for arbitrary collection of two-qubit entangled states in the network. Now, let each source distribute arbitrary two qubit state. Also, let each of the two extreme parties can perform incompatible measurements. When all these resources are available, \textit{will non $n$-locality be detected for any set of incompatible measurements performed by extreme parties?} We provide negative response to this query.
	\begin{theorem}\label{thinc01}
		In any standard linear $n$-local network, non $n$-locality cannot be detected even if all the extreme parties perform incompatible measurements.
	\end{theorem}
		\textit{Proof:} See Appendix.\ref{appnc}\\
	It is observed that in case each of the extreme parties perform Pauli measurements along $X$ and $Z$ directions, resulting network correlations never violate $n$-local inequality(Eq.(\ref{ineqb})) irrespective of the states distributed by the sources(see Appendix.\ref{appnc}). This result is in contrast to the recent findings corresponding to Bell-CHSH scenario where nonlocality can always be exploited in case both parties perform any set of incompatible projective measurements\cite{ujjwal}. 
	
	Theorem.\ref{thinc1} points out that to detect non $n$-local correlations, all the parties need not perform incompatible measurements. Theorem.\ref{thinc01} implies that for any set of $n$ arbitrary two-qubit states, non $n$-locality cannot always be detected even if both the extreme parties perform incompatible measurements. However, such detection relies on violation of $n$-local inequality which is just a sufficient criterion\cite{Bran}. So it becomes pertinent to inquire whether non $n$-local correlations can be generated(absence of HVM) if none of the parties perform incompatible measurement. It turns out that non $n$-locality can still be generated for any $n$$\geq$$3.$ However for $n$$=$$2,$ we get a negative response.
	\begin{theorem}\label{theo2}
		In any standard linear $n$-local network with $n$$\geq$$3,$ non $n$-locality can be generated even when all the parties perform compatible measurements. However, for $n$$=$$2,$ correlations admit bilocal model for the same measurement context.
	\end{theorem} 
	\textit{Proof:} See Appendix.\ref{appnd}.\\
	In Bell scenario, incompatibility of measurements is necessary as correlations admit LHV model even when one of the parties perform incompatible measurements\cite{Fin,Wol}. Theorem.\ref{theo2} justifies in-equivalence between Bell scenario and network scenario in context of generating non-classical correlations.\\
	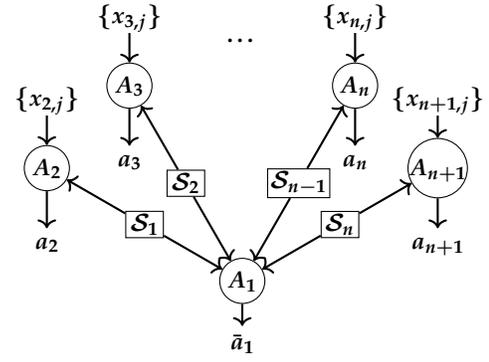
\begin{figure}
		\begin{tikzpicture}[
			node distance=1.4cm and 1.6cm,
			every node/.style={font=\boldmath\bfseries\small},
			meas/.style={
				draw, circle, minimum size=0.6cm,
				inner sep=0pt, align=center,
				text height=1.2ex, text depth=0ex
			},
			source/.style={
				draw, rectangle, minimum width=0.5cm, minimum height=0.3cm,
				inner sep=1pt, align=center
			},
			arr/.style={->, thick}
			]
			
			\node[meas] (A1) at (0,0) {$A_1$};
			
			\node[meas] (A2) at (150:3cm) {$A_2$};
			\node[meas] (A3) at (120:3cm) {$A_3$};
			
			\node[meas] (An) at (60:3cm) {$A_n$};
			\node[meas] (Anp1) at (30:3cm) {$A_{n+1}$};
			
			\node at (90:3.2cm) {\dots};
			
			\node[source] (S1) at ($(A1)!0.5!(A2)$) {$\mathcal{S}_1$};
			\node[source] (S2) at ($(A1)!0.5!(A3)$) {$\mathcal{S}_2$};
			\node[source] (Sn) at ($(A1)!0.5!(An)$) {$\mathcal{S}_{n-1}$};
			\node[source] (Snp1) at ($(A1)!0.5!(Anp1)$) {$\mathcal{S}_n$};
			
			\foreach \X in {S1, S2, Sn, Snp1} {
				\draw[arr] (\X) -- (A1);
			}
			\draw[arr] (S1) -- (A2);
			\draw[arr] (S2) -- (A3);
			\draw[arr] (Sn) -- (An);
			\draw[arr] (Snp1) -- (Anp1);
			
			\foreach \i/\A in {2/A2, 3/A3, n/An, {n+1}/Anp1} {
				\node at ($(\A)+(0,0.9)$) {$\{x_{\i,j}\}$};
				\draw[arr] ($(\A)+(0,0.7)$) -- (\A);
				\draw[arr] (\A) -- ++(0,-0.8) node[below] {$a_{\i}$};
			}
			
			\draw[arr] (A1) -- ++(0,-0.6) node[below] {$\bar{a}_1$};
		\end{tikzpicture}
		\caption{Schematic Diagram of $n$-local star network}
		\label{nlocal2}
	\end{figure}
	All above theorems provide an idea about role of linear network topology to generate quantum non-classicality with lesser resource(in terms of measurement incompatibility \cite{Hei,Gue,Sky}). For gaining better intuition about the broader implications of network architecture in exploiting such non-classicality, we next consider a non-linear network, specifically $n$-local star network \cite{Tava}. It was pointed out in \cite{Taav,rece} that star topology offers advantage over linear one in context of generating non $n$-local correlations. In such a network(see Fig.\ref{nlocal2}), there is one central party with single input, specifically performing $n$-partite GHZ basis measurement whereas each of $n$ extreme parties chooses from a set of two dichotomic measurements \cite{Tava}. So we need to explore whether for detecting non $n$-locality, all or some of the extreme parties need to choose from a set of incompatible measurements. Following $n$-local inequality's(Eq.\ref{ineqsn}) violation is sufficient to detect non $n$-locality\cite{Tava}:
	\begin{eqnarray}\label{ineqsn}
		&& \frac{1}{ 2^{n-2}}\sum_{i=1}^{2^{n-1}}|J_{i}|^{\frac{1}{n}}\leq 1,\,  \textmd{where}\\
		&& J_i=\frac{1}{2^n}\sum_{x_2,...,x_{n+1}} (-1)^{s_i (x_2,...,x_{n+1})}\langle D_{ (1)}^{ (i)}D_{x_2}^{ (2)}...D_{x_{n+1}}^{ (n+1)}\rangle\nonumber\\
		&&\langle D_{ (1)}^{ (i)}D_{x_2}^{ (2)}...D_{x_{n+1}}^{ (n+1)}\rangle=\sum_{\mathcal{D}_2} (-1)^{\tilde{a}_1^{ (i)}+a_2+...
			+a_{n+1}}N_2,\nonumber\\
		&& \textmd{\small{where}}\,N_2=\small{p (\overline{\textbf{a}}_1,
			a_2,...,a_{n+1}
			|x_2,...,x_{n+1})}\,\textmd{\small{and}}\nonumber\\
		&& \mathcal{D}_2=\{a_{11},....,a_{12^n},a_2,....
		,a_{n+1}\}\nonumber\\
	\end{eqnarray}
	In Eq.(\ref{ineqsn}) $\forall i$$=1,...,2^{n-1},$ $\tilde{a}_1^{ (i)}$ denotes an output bit obtained by classical post-processing of the raw output string $\overline{a_1}$$=$$ (a_{11},....,a_{1n})$ of $A_1.$ $\forall i,\, s_i$ are functions of the input variables $x_{2,j},...,x_{n+1,j}$ of the extreme parties \cite{Tava}. Each $s_i$ is function of even number of input variables. \\
	\par Consider $n$-local star network where each of the sources distribute two-qubit entangled state. Let the central party $A_1$ perform $n$-partite GHZ-basis measurement. Each of $n$ extreme parties $A_2,...,A_{n+1}$ chooses from a set of two dichotomic measurements. Interestingly, non $n$-locality can be obtained when only one of these extreme parties performs from a set of incompatible measurements. We provide our observation in this context.
	\begin{theorem}\label{theo3}
	In any standard $n$-local star network, non $n$-locality can be detected even if all the parties do not perform incompatible measurements.	
\end{theorem}

\textit{Proof:} See Appendix for the proof.\ref{appne}.	Detailed characterization of quantum states along with incompatible settings for one extreme party and compatible settings for remaining are provided therein .\\
For numerical illustration, let us consider $3$ Werner states with $v_1$$=$$v_2$$=$$v_3$$=$$0.93$ in trilocal star network. Let $x_{i,j}$$=$$\{\eta_{i-1}(\sigma_1+(-1)^j\sigma_2)\}_{j=0}^1$ for $i$$=$$2,3,4$ with $\eta_1$$=$$0.672,$ $\eta_2$$=$$0.5$ and $\eta_3$$=$$0.488.$ For this set-up, $V_{\small{n-star}}^{\eta_1,\eta_2,\eta_3}$$=$$1.018.$ Non-trilocal correlations are thus detected. \\
Now consider Werner state with $v$$=$$0.74.$ Let identical copies of this state be used in a linear $4$-local network and also separately in $4$-local star network. In each network, let only one of the extreme parties perform incompatible measurement whereas remaining extreme parties perform compatible measurement. For some measurement parameters non $4$-locality is detected(via violation of $4$-local inequality) in $4$-local star network. But for optimal projective measurement contexts, non $4$-locality is not detected in linear network(see Appendix.\ref{appnf} for details). This example points out possibility of detecting stronger non $n$-local correlations in star topology when only one party performs incompatible measurements. Such advantage, in terms of exploiting quantumness, offered by star network over linear one appears to stem from the structure inherent in the star-shaped architecture. Here central party, getting access to more than $2$ qubits, can perform measurement in genuinely entangled basis in contrast to linear set-up. \\
Though star topology surpasses over linear one, yet result analogous to that provided by Theorem.\ref{thinc01} exists for $n$-local star network:
\begin{theorem}\label{theo30}
	In any standard $n$-local star network, non $n$-locality cannot be detected even if all the extreme parties perform incompatible measurements.
\end{theorem}	
\textit{Proof:} See Appendix.\ref{appng} for details.\\
Above theorem points out limitations(in terms of measurement settings) over detection of non $n$-local correlations even if network involves maximally entangled two-qubit states.
\par In standard $n$-local star network, correlations are $n$-local if those satisfy $n$-local constraint and can be factorized:
\begin{eqnarray}\label{tr2n}
	&&\small{P (\overline{a}_1,a_2,a_3...,a_{n+1}|x_2,x_3,...,,x_{n+1})}=\nonumber\\
	&&\sum_{\lambda_1\in\Lambda_1}...\sum_{\lambda_n\in\Lambda_n} \mu(\lambda_1,...,\lambda_n)\mathcal{P}_2\,\textmd{\small{with}}\nonumber\nonumber\\
	&&\mathcal{P}_2=P (\overline{a}_1|\lambda_1,...,\lambda_n)\Pi_{i=1}^{n} p (a_{i+1}|x_{i+1},\lambda_i)\nonumber\\
	&&
\end{eqnarray}
Correlations are explicable in above form(Eq.(\ref{tr2n})) if all parties perform compatible measurements. Next theorem justifies this claim. 
\begin{theorem}\label{theo4}
	In any standard $n$-local star network, correlations admit $n$-local hidden variable model if all extreme parties perform compatible measurements.
\end{theorem} 
\textit{Proof:}The proof is similar to the proof of second part of Theorem.\ref{theo2}.\\
Above result points out impossibility to generate non $n$-local correlations if the entire measurement context, corresponding to star topology, is devoid of compatible measurements. This in contrast to our findings in linear topology with at least three sources. Such an observation is quite intuitive as stronger correlations are obtained for standard network having star topology. Even more stronger quantum correlations are obtained in non standard network characterized by absence of any source with local behavior. As pointed out in \cite{Poz}, genuine form of network nonlocality can be obtained only in any non standard quantum network. In this context, it becomes necessary to analyze precise role of measurement incompatibility as a resource to generate genuine network correlations. We provide related observations in next theorem.
\begin{theorem}\label{corr1}
	To generate fully network nonlocal correlations in any non standard $n$-local network with arbitrary quantum sources, all the edge parties must perform incompatible measurements.
\end{theorem}
\textit{Proof:} See Appendix.\ref{appnh}\\
Thus, unlike standard network, genuine network non-classicality cannot be exhibited even if only one edge party performs compatible measurements.\\
\textit{Discussions:} Considering the aspect of incompatibility in quantum measurements, our study justifies a direction of operational in-equivalence between Bell nonlocality and non $n$-locality in any standard $n$-local network. Precisely, our observations point out that incompatibility in measurement settings of all parties is not a mandate to generate non $n$-local correlations in any such network. From such perspective, a more systematic study is needed to explore the intrinsic factors, pertaining to any quantum network, that are responsible for such in-equivalence. \\
\par  Apart from implications about lesser need of resource, our findings also prescribe the compatible measurement contexts along with characterization of the quantum sources for which non $n$-locality can be witnessed. Also, we got the idea about the minimum resource(measurement incompatibility) requirement in this context. It turns out that in linear bilocal network, tripartite correlations admit bilocal HV model in absence of measurement incompatibility. However, measurement incompatibility is not required if we increase the length of the linear chain involving at least $3$ independent sources. But for star topology the minimum requirement of measurement incompatibility(on one party) persists for any number of sources. Such a difference between linear and star topology aligns with existing claim that one can obtain stronger quantum correlations in the latter. A comprehensive study on the difference between different topology in standard $n$-local network in terms of measurement incompatibility will be interesting. Also study on relationship between different recent notions of measurement incompatibility(see\cite{recent} and references therein) and network nonlocality is a potential direction of future research                     
	
	\section{Appendix.A}\label{appna}
	Consider a set of two noisy Von Neumann equatorial measurements $\{M_1,M_2\}$ where:
	\begin{eqnarray}\label{app11}
		M_j&=&\eta\vec{n}_j\cdot \vec{\sigma}\,\,	\textmd{\small{with}}\eta\in[0,1]\,\,||\vec{n}_j||\leq 1\\
		\textmd{\small{and}}&& \vec{n}_j=(\sin(t),0,(-1)^j\cos(t))\,\,j=0,1\nonumber
	\end{eqnarray}
	Note that for $t$$=$$\frac{z\pi}{2}$($z$ being any integer), $M_0,M_1$ are always compatible. So we try to find out range of $\eta$ for which above set is compatible considering $t$$\neq$$\frac{z\pi}{2}.$ 
	Above set of measurements is compatible if and only if \cite{Bus}:
	\begin{eqnarray}\label{app12}
		\eta(||\vec{n}_0+\vec{n}_1||+||\vec{n}_0-\vec{n}_1||)\leq 2\nonumber\\
		\Rightarrow \eta\leq \frac{1}{|\cos(t)|+|\sin(t)|}
	\end{eqnarray}
	Following similar strategy, one can see that the set $\{\eta \vec{n}_0.\vec{\sigma},\eta \vec{n}_1.\vec{\sigma}\},$ with $\vec{n}_j$$=$$(\cos(t),(-1)^j\sin(t),0),$ is compatible if and only if Eq.(\ref{app12}) is satisfied.\\
	Let us now use above criterion to show that $\{x_{1,0},x_{1,1}\}$ and $\{x_{3,0},x_{3,1}\}$ that we have used in our example in the main text for showing BRGP inequality violation, are set of incompatible and compatible measurements respectively.\\
	Let $t$$=$$\frac{\pi}{4}$ and $\vec{n}_j$$=$$(\sin(t),0,(-1)^j\cos(t)).$  Then by Eq.(\ref{app12}), we have $\{M_0,M_1\}$$=$$\{\eta(\frac{\sigma_3\pm\sigma_1}{\sqrt{2}})\}$ to be compatible iff $\eta$$\leq$$\frac{1}{\sqrt{2}}.$
	In the example, we have chosen $x_{1,i}$$=$$0.664(\sigma_3+(-1)^i\sigma_1)$$=$$ \frac{\eta_1}{\sqrt{2}}(\sigma_3+(-1)^i\sigma_1)$ with $\eta_1$$=$$0.939$$>$$\frac{1}{\sqrt{2}}.$ Hence $\{x_{1,0},x_{1,1}\}$ is incompatible.
	Again $x_{3,i}$$=$$0.494(\sigma_3+(-1)^i\sigma_1)$$=$$ \frac{\eta_2}{\sqrt{2}}(\sigma_3+(-1)^i\sigma_1)$ with $\eta_2$$=$$0.699$$<$$\frac{1}{\sqrt{2}}.$ Hence $\{x_{3,0},x_{3,1}\}$ is compatible.
	\section{Appendix.B}\label{appnb}
	\subsection*{Discussion on existing $n$-local inequality(Eq.(\ref{ineqb}))}
	For $n$$=$$2,$ Eq.(\ref{ineqb}) gives the BRGP inequality.\\
Upper bound $B_{\small{n-lin}}$(say) of Eq.(\ref{ineqb}) is of the form\cite{Gis}:
\begin{equation}\label{blin}
	B_{\small{n-lin}}=\sqrt{\Pi_{i=1}^n E_{i1}+\Pi_{i=1}^n E_{i2}}
\end{equation}
with $E_{i1}$$\geq$$E_{i2}$ denoting the two largest singular values of $\rho_i^{'}s$ correlation tensor $\forall i$$=$$1,2,...,n$.\\
The upper bound $B_{\small{n-lin}}$ of Eq.(\ref{ineqb}) is achievable when \cite{Gis}:
\begin{itemize}
	\item $A_1$ performs $x_{1,j}$ where:
	\begin{eqnarray}\label{ainc1}
		x_{1,j}&=&\{(\cos (r)\sigma_3+(-1)^j\sin (r)\sigma_1)\}_{j=0}^1,\\
		\textmd{\small{where}}\,r&=&\arcsin\sqrt{\frac{\Pi_{i=1}^nE_{i1}}{\Pi_{i=1}^nE_{i1}+\Pi_{i=1}^nE_{i2}}}
	\end{eqnarray}
	\item $A_{n+1}$ performs $x_{n+1,j}$ such that $x_{n+1,j}$$=$$x_{1,j},$ for $j$$=$$0,1.$  
	\item Each of $A_2,A_3,...,A_n$ performs Bell basis measurement: 
	\begin{eqnarray}\label{bellbasis}
		\mathbf{B}&=&\{P_{|\phi^+\rangle},P_{|\phi^-\rangle},P_{|\psi^+\rangle},P_{|\psi^-\rangle}\},
	\end{eqnarray}
	$P_{|\phi^+\rangle},P_{|\phi^-\rangle},P_{|\psi^+\rangle},P_{|\psi^-\rangle}$ denoting projectors along Bell states $|\phi^+\rangle,|\phi^-\rangle,|\psi^+\rangle,|\psi^-\rangle$ respectively.
\end{itemize}
If violation is observed then:
\begin{equation}\label{maxvio}
	1<B_{\small{n-lin}}\leq \sqrt{2}
\end{equation}
Now, we give the condition over the set of $n$ two-qubit states used in the network for which Theorem.\ref{thinc1} holds.
\subsection*{Conditions over states} Each of $n$ sources $\mathcal{S}_i$ is generating two-qubit state $\rho_i$. We consider those $\rho_1,\rho_2,...,\rho_n$ for which the following criterion holds:
\begin{eqnarray}\label{theoe01}
	\Pi_{i=1}^n E_{i1}+\Pi_{i=1}^n E_{i2}&>& (\sqrt{\Pi_{i=1}^n E_{i1}}+\sqrt{\Pi_{i=1}^n E_{i2}})^{\frac{2}{3}}
\end{eqnarray}
To prove the theorem, we now only need to prove existence of suitable measurement context with only one extreme party performing some fixed incompatible measurements for which violation is observed in the network involving states satisfying above mentioned criteria.\\
Next, we give the fixed incompatible measurement settings for one of the extreme parties.
\subsection*{Fixed Incompatible Set of Measurement for One Extreme Party} 
W.L.O.G., we consider fixed incompatible measurement settings for party $A_1.$ We consider the following set of two dichotomic measurement settings:
\begin{eqnarray}\label{inc4}
	x_{1,j}&=&\{\eta_1(\sin(r) \sigma_3+(-1)^j\cos(r) \sigma_1)\}_{j=0}^1,\\
	r&=&\arcsin\sqrt{\frac{\Pi_{i=1}^nE_{i1}}{\Pi_{i=1}^nE_{i1}+\Pi_{i=1}^nE_{i2}}}\nonumber\\
	1&\geq&\eta_1>	\frac{\cos(r)+\sin(r)}{\Pi_{i=1}^n E_{i1}+\Pi_{i=1}^n E_{i2}}
\end{eqnarray}
In Eq.(\ref{inc4}), $E_{i1}$$\geq$$E_{i2}$ are the two largest singular values of $\rho_i^{'}s$ correlation tensor $\forall i$$=$$1,2,...,n$. \\
By given condition(Eq.(\ref{theoe01})) over the states and the expression of argument $r$(in Eq.\ref{inc4}), we get:
\begin{eqnarray}\label{theoe1}
	\Pi_{i=1}^n E_{i1}+\Pi_{i=1}^n E_{i2}&>& \cos(r)+\sin(r)\nonumber\\
	\Rightarrow \frac{\cos(r)+\sin(r)}{\Pi_{i=1}^n E_{i1}+\Pi_{i=1}^n E_{i2}}&< &1.
\end{eqnarray}

Hence, $(\frac{\cos(r)+\sin(r)}{\Pi_{i=1}^n E_{i1}+\Pi_{i=1}^n E_{i2}},1]$ is a valid range of noise parameter $\eta_1$.\\
Clearly, $r$$\neq$$z$(integer) in Eq.(\ref{inc4}). Also this set of measurements violates criterion of compatibility(Eq.(\ref{app12}). Hence Eq.(\ref{inc4}) represents incompatible set of measurements.
\subsection*{Proof of Existence of Compatible Measurements By $A_{n+1}$}
As per our requirement, we need to find out compatible measurements for $A_{n+1}$ such that violation of Eq.(\ref{ineqb}) is observed in the network.\\
$A_1$ performs from the set of $2$ incompatible measurements given by Eq.(\ref{inc4}).
Let each of the central parties $A_2,A_3,..,A_n$ performs only Bell basis measurement(Eq.(\ref{bellbasis})). \\
We now need to find out a set of two compatible measurements for the remaining party $A_{n+1}$ such that corresponding $n+1$-partite correlations violate $n$-local inequality(Eq(\ref{ineqb})).\\
Let $A_{n+1}$ perform $x_{n+1,0},x_{n+1,1}$ such that:
\begin{eqnarray}\label{ainc2}
	x_{n+1,j}&=&\{\eta_2(\sin (r) \sigma_3+(-1)^j\cos (r) \sigma_1)\}_{j=0}^1,\,\eta_2\in [0,1]\nonumber\\
	r&=&\arcsin\sqrt{\frac{\Pi_{i=1}^nE_{i1}}{\Pi_{i=1}^nE_{i1}+\Pi_{i=1}^nE_{i2}}}
\end{eqnarray}
For these measurement settings, upper bound($B_{\small{n-lin}}$) of $n$-local inequality(Eq.(\ref{ineqb})) gets modified to  $B_{\small{n-lin}}^{(\eta_1,\eta_2)}$(say). This modified bound is given by \cite{Bis}:
\begin{eqnarray}\label{ainc3}
	B_{\small{n-lin}}^{(\eta_1,\eta_2)}&=&\sqrt{\eta_1\cdot\eta_2(\Pi_{i=1}^n E_{i1}+\Pi_{i=1}^n E_{i2})}\nonumber\\
	&=&\sqrt{\eta_1\cdot\eta_2}\cdot B_{\small{n-lin}}
\end{eqnarray}
Now, the set of measurement settings provided by Eq.(\ref{ainc2}) is compatible if:
\begin{equation}\label{ainc3i}
	\eta_2\leq \frac{1}{\cos(r)+\sin(r)}\,\,(\textmd{\small{Here }}\cos(r),\sin(r)>0).
\end{equation}
As non $n$-locality is detected in the network via violation of Eq.(\ref{ineqb}), it follows from Eq.(\ref{ainc3}) that $\eta_2$ must satisfy the following condition:
\begin{equation}\label{aincs5}
	\eta_2>\frac{1}{\eta_1\cdot B_{\small{n-lin}}^2}
\end{equation} 
Any $\eta_2$ satisfying both Eq.(\ref{ainc3i}) and Eq.(\ref{aincs5}) will meet up our requirements. \\
Now, by given conditions over $\eta_1$(Eq.\ref{inc4}), we have:
\begin{eqnarray}
	\eta_1&>&\frac{\cos(r)+\sin(r)}{B_{\small{n-lin}}^2}\nonumber\\
	\Rightarrow	\frac{1}{\eta_1\cdot B_{\small{n-lin}}^2}&<&\frac{1}{\cos(r)+\sin(r)}
\end{eqnarray}
Also, by Eqs.(\ref{inc4},\ref{aincs5}), $	\frac{1}{\eta_1\cdot B_{\small{n-lin}}^2}$$>$$0.$ \\
Hence, $(\frac{1}{\eta_1\cdot B_{\small{n-lin}}^2},\frac{1}{\cos(r)+\sin(r)}]$ is an interval of real numbers such that 
$(\frac{1}{\eta_1\cdot B_{\small{n-lin}}^2},\frac{1}{\cos(r)+\sin(r)}]$$\subset$$[0,\frac{1}{\cos(r)+\sin(r)}].$\\
Let $r^{'}$$\in$$(\frac{1}{\eta_1\cdot B_{\small{n-lin}}^2},\frac{1}{\cos(r)+\sin(r)}].$\\
Setting $\eta_2$$=$$r^{'}$ in Eq.(\ref{ainc2}) will suffice for our purpose.\\
This completes our search of compatible measurement set for $A_{n+1}.$\\
Theorem is thus proved$\blacksquare$
	\section{Appendix.C}\label{appnc}
	Let each of $n$ independent sources $\mathcal{S}_1,...,\mathcal{S}_n$ distribute arbitrary two-qubit state $\rho_1,...,\rho_n$ respectively.
	Using Bloch parameters $\rho_i$ can be written as:
	\begin{eqnarray}\label{st4}
		\small{\rho}_i&=&\small{\frac{1}{4}(\mathbb{I}_{2\times2}+\vec{u}_i.\vec{\sigma}\otimes \mathbb{I}_2+\mathbb{I}_2\otimes \vec{v}_i.\vec{\sigma}}+\nonumber\\
		&&\small{\sum_{l_1,l_2=1}^{3}T_{i  l_1 l_2}\sigma_{l_1}\otimes\sigma_{l_2})},\,\forall i=1,...,n.
	\end{eqnarray}
	$\forall i$$=$$1,2,...,n,$ $\vec{u}_i,$ $\vec{v}_i$$\in$$\mathbb{R}^3$ represent the local bloch vectors and
	$(T_{ijk})_{3\times3}$ denotes the correlation tensor $\mathcal{T}_i$(real matrix) of $\rho_i$ where 
	$t_{ijk}$$=$$\textmd{Tr}[\rho_i\cdot\sigma_{j}\otimes\sigma_{k}].$ \\
	$\mathcal{T}_i$ can be diagonalized via suitable local unitary operations\cite{Luo}:
	\begin{equation}\label{st41}
		\small{\rho}_i^{'}=\small{\frac{1}{4}(\mathbb{I}_{2\times2}+\vec{a}_i.\vec{\sigma}\otimes \mathbb{I}_2+\mathbb{I}_2\otimes \vec{b}_i.\vec{\sigma}+\sum_{j=1}^{3}E_{ij}\sigma_{j}\otimes\sigma_{j})},\,\forall i=1,...,n
	\end{equation}
	Here the correlation tensor is given by $\textbf{E}_i$$=$$\textmd{diag}(E_{i1},E_{i2},E_{i3}).$ $E_{i1},E_{i,2},E_{i,3}$ are the eigenvalues of $\sqrt{\mathcal{T}_i^T\mathcal{T}_i},$ i.e., singular values of $\mathcal{T}_i$ arranged in descending order of magnitude, i.e., $E_{i1}$$\geq$$E_{i2}$$\geq$$E_{i3}.$
	According to the given measurement settings:
	\begin{itemize}
		\item $\{x_{1,0},x_{1,1}\}$$=$$\{\sigma_1,\sigma_3\}$
		\item $\{x_{n+1,0},x_{n+1,1}\}$$=$$\{\sigma_1,\sigma_3\}$
		\item Each of $A_2,...,A_n$ performing Bell basis measurement.
	\end{itemize}
	W.L.O.G., let us consider the following specifications:
	\begin{eqnarray*}
		x_{1,0}&=&\sigma_1; x_{1,1}=\sigma_3; \\
		x_{n+1,0}&=&\sigma_1; x_{n+1,1}=\sigma_3; 
	\end{eqnarray*}
	For above measurement settings, the correlator terms $I_n,J_n$ in $n$-local inequality(Eq.(\ref{ineqb})) take the form:
	\begin{eqnarray}\label{appc1}
		I_n&=&\frac{1}{4}\textmd{Tr}[(\sigma_1+\sigma_3)\otimes(\sigma_3)^{\otimes n-1}\otimes (\sigma_1+\sigma_3)\cdot \otimes_{i=1}^n\rho_i]\nonumber\\
		I_n&=&\frac{1}{4}\textmd{Tr}[(\sigma_1-\sigma_3)\otimes(\sigma_1)^{\otimes n-1}\otimes (\sigma_1-\sigma_3)\cdot \otimes_{i=1}^n\rho_i]\nonumber\\
		&&
	\end{eqnarray}
	Simplifying above form of correlators(Eq.(\ref{appc1})), L.H.S. of Eq.(\ref{ineqb}) turns out to be:
	\begin{eqnarray*}
		\sqrt{|I_n|}+\sqrt{|J_n|}&=&\frac{1}{2}(\sqrt{\Pi_{i=1}^n E_{i1}}+\sqrt{\Pi_{i=1}^n E_{i3}})\\
		&\leq& 1
	\end{eqnarray*}
	Hence, $n$-local inequality(Eq.(\ref{ineqb})) is not violated for any $\rho_1,\rho_2,...,\rho_n.$ 
	\section{Appendix.D}\label{appnd}
	\textit{Proof of Theorem.\ref{theo2}:} We will first prove second part of the theorem. Then we will prove the first part.\\
	Let us consider a linear bilocal network. Let both the extreme parties $A_1$ and $A_3$ choose from a set of 2 dichotomic compatible measurements: $\{x_{1,0},x_{1,1}\}$ and $\{x_{3,0},x_{3,1}\}$ respectively. \\
	For $i$$=$$1,3,$ let $a_i$ denote the binary valued outputs corresponding to input $x_{i,j},$ $\forall j$$=$$0,1.$  \\
	Let central party $A_2$ perform single measurement $x_2$(say).\\
	$\forall i$$=$$1,3$ and $j$$=$$0,1$ let $M_{i}^{a_i|x_{i,j}}$ denote POVM elements corresponding to $2$ outputs($a_i$$\in$$0,1$) of $x_{i,j}$\\
	As $\{x_{1,0},x_{1,1}\}$ is a set of dichotomic compatible measurements, so there exists a parent POVM $\{G_{1,\lambda}\}_{\lambda}$ such that:
	\begin{eqnarray}\label{meas1}
		M_{1}^{a_1|x_{1,j}}&=&\sum_{\lambda} p(a_1|x_{1,j},\lambda) G_{1,\lambda},\,\forall j,a_1\in\{0,1\}\\\
	\end{eqnarray}
	Similarly, there exists a parent POVM $\{G_{3,\nu}\}_{\nu}$ for $A_3^{'}$s input set($\{x_{3,0},x_{3,1}\}$) such that: 
	\begin{eqnarray}\label{meas2}
		M_{3}^{a_3|x_{3,j}}&=&\sum_{\nu} p(a_3|x_{3,j},\nu) G_{3,\nu},\,\forall j,a_3=0,1\\
	\end{eqnarray}
	For ease of writing, let $(a_i,x_i)$ denote any (output,input) pair of $A_i$($i$$=$$1,3$). Then $x_i$$\in$$\{x_{i,0},x_{i,1}\}.$\\
	Let $\bar{a}_2$ denote two bit output string of $A_2:$ $\bar{a}_2$$=$$(a_{21},a_{22})$ with $a_{21},a_{22}$$\in$$\{0,1\}.$ Let $M_{\bar{a}_2}$ denote POVM element corresponding to any output bit string $\bar{a}_2$ of $A_2$.\\
	Let $\mathcal{S}_1,\mathcal{S}_2$ distribute arbitrary two-qubit state $\rho_1,\rho_2$ respectively. With these notations, we can write any probability term $P(a_1,\bar{b},a_3|x_1,x_3)$ as:
	\begin{eqnarray}\label{meas3}
		P(a_1,\bar{a}_2,a_3|x_1,x_3)&=&\textmd{Tr}[(M_{1}^{a_1|x_1}\otimes M_{\bar{a}_2}\otimes M_{3}^{a_3|x_3}) \cdot \rho_1\otimes\rho_2]\nonumber\\
		&=&\sum_{\lambda,\nu}  P(a_1|x_{1},\lambda)\cdot P(a_3|x_{3},\nu) \textmd{Tr}[(G_{1,\lambda}\otimes \nonumber\\
		&&M_{\bar{a}_2} \otimes G_{3,\nu})\cdot \rho_1\otimes\rho_2]\nonumber\\
		&=&\sum_{\lambda,\nu}  P(a_1|x_{1},\lambda)\cdot P(a_3|x_{3},\nu) P(\bar{a}_2,\lambda,\nu)\nonumber\\
		&=&\sum_{\lambda,\nu}  P(a_1|x_{1},\lambda)\cdot P(a_3|x_{3},\nu)\nonumber\\
		&&\cdot P(\bar{a}_2|\lambda,\nu)\cdot P(\lambda,\nu).\nonumber\\
		&&
	\end{eqnarray}
	Also, $\lambda,\nu$ characterize measurements of two spatially separated parties($A_1,A_3$). Hence they are independent:
	\begin{equation}\label{meas4}
		P(\lambda,\nu)=P(\lambda)P(\nu)
	\end{equation}
	So here any measurement probability term satisfies both Eq.(\ref{meas3}) and Eq.(\ref{meas4}). Hence they are bilocal in nature.\\
	This proves that when both the extreme parties choose from a set of 2 dichotomic compatible measurements and central party performs single measurement, measurement probability terms admit a bilocal model. This proves second part of the theorem.\\
	Let us now prove the first part of the theorem.\\
	Let us consider a linear network involving four parties arranged sequentially, such that each pair of neighbouring parties shares a common source. As discussed in the main text, in this configuration(linear chain network) the two extreme parties perform two measurements each, while the intermediate parties perform only a single measurement(see Fig.\ref{nlocal1} ). However, due to the constraint imposed by the relevant theorem, all measurements are assumed to be compatible. Consequently, each party effectively performs a single measurement on the particle it receives. In each run of the experiment, the parties record their respective outcomes, denoted by $a_1$, $a_2$, $a_3$, $a_4$. Repeating the experiment many times allows the parties to estimate correlations among the outcomes and thus determine the joint probability distribution$P(a_1, a_2,a_3,a_4)$. As shown in \cite{Bran} and \cite{Fri}, such a correlation $P(a_1, a_2,a_3,a_4)$ can be interpreted within a conventional bipartite framework as a no-signalling box, accompanied by input distributions $P(a_1)$ and $P(a_4)$. Importantly, the overall correlation is said to be network-local if and only if the associated no-signalling box $P(a_2,a_3|a_1,a_4)$ is Bell-local. Since it is known that there exist entangled states for which the correlation $P(a_2,a_3 | a_1,a_4)$ violates a Bell inequality(\cite{Bru}), the existence of network nonlocal correlations is thereby guaranteed in this four-party linear network even when all measurements are compatible.\\
	Similarly, when all parties in a five-partite linear network perform compatible measurements, it has been shown in \cite{Fri} that this scenario reduces to the three-partite (bilocal) linear network configuration. In this setting, a five-partite correlation of the form $P(a_1,a_2,a_3,a_4,a_5)$ is considered. It turns out that the correlation $P(a_2,a_3,a_4,a_1,a_5)$ is network local if and only if the associated conditional distribution $P(a_2,a_3,a_4 \mid a_1,a_5)$ is network local in the bilocal sense \cite{Fri}. Since it is known that there exist entangled quantum states which lead to a violation of the BRGP inequality \cite{Bran} in the distribution $P(a_2,a_3,a_4 \mid a_1,a_5)$, the presence of network nonlocal correlations is thereby guaranteed in the five-partite linear $5$-local network even under the constraint that all measurements are compatible.
	This reasoning can be generalized to any $n+1$-partite linear network($n \geq 3$). Specifically, a $n+1$-partite correlation $P(a_1, a_2, \dots, a_{n+1})$ is network local if and only if the associated conditional distribution $P(a_2, a_3, \dots, a_{n} \mid a_1, a_{n+1})$ is network local. As established in the literature\cite{Kau}, there exist quantum correlations for which this conditional distribution exhibits network nonlocality. Hence, network nonlocality is also present in the full distribution $P(a_1, a_2, \dots, a_n,a_{n+1})$.\\
	Therefore, we conclude that network nonlocality is guaranteed in any $n+1$-partite linear network configuration for $n$$\geq$$3$, even when all parties perform only compatible measurements.
	\section{Appendix.E}\label{appne}	
	\subsection*{Discussion on Existing $n$-Local Inequality(Eq.(\ref{ineqsn})) in Star Network \cite{Tava}}
	Recalling existing $n$-local inequality(Eq.(\ref{ineqsn})) we have:
	\begin{eqnarray*}
		&& \frac{1}{ 2^{n-2}}\sum_{i=1}^{2^{n-1}}|J_{i}|^{\frac{1}{n}}\leq 1,\,  \textmd{where}\\
		&& J_i=\frac{1}{2^n}\sum_{x_2,...,x_{n+1}} (-1)^{s_i (x_2,...,x_{n+1})}\langle D_{ (1)}^{ (i)}D_{x_2}^{ (2)}...D_{x_{n+1}}^{ (n+1)}\rangle\nonumber\\
		&&\langle D_{ (1)}^{ (i)}D_{x_2}^{ (2)}...D_{x_{n+1}}^{ (n+1)}\rangle=\sum_{\mathcal{D}_2} (-1)^{\tilde{a}_1^{ (i)}+a_2+...
			+a_{n+1}}N_2,\nonumber\\
		&& \textmd{\small{where}}\,N_2=\small{p (\overline{\textbf{a}}_1,
			a_2,...,a_{n+1}
			|x_2,...,x_{n+1})}\,\,\textmd{\small{and}}\nonumber\\
		&& \mathcal{D}_2=\{a_{11},....,a_{12^n},a_2,....
		,a_{n+1}\}\nonumber\\
	\end{eqnarray*}
	$\forall i$$=1,...,2^{n-1},$ $\tilde{a}_1^{ (i)}$ denotes an output bit obtained by classical post-processing of the raw output string $\overline{a_1}$$=$$ (a_{11},....,a_{1n})$ of $A_1.$\cite{Tava}. Also $\forall i,\,s_i$ is function of even number of input variables\cite{Tava}.
	There are total $2^{n-1}$ correlator terms $J_i.$\\
	Let each of the sources $S_1,...,S_n$ distribute an arbitrary two qubit state $\rho_i$.\\
	As in \cite{Tava}, we consider the following measurement context:
	\begin{itemize}
		\item $A_1$ performs $n$-partite GHZ basis measurement.
		\item $\forall i$$=$$2,3,...,n$$+$$1,$ $A_i$ performs $x_{i,j}$ where:
		\begin{eqnarray}\label{binc1}
			x_{i,j}&=&\{(\cos (t)\sigma_1+(-1)^j\sin (t)\sigma_2)\}_{j=0}^1,\\
			\textmd{\small{where}}\,t&=&\arcsin\sqrt{\frac{\Pi_{i=1}^n (E_{i2})^{\frac{2}{n^2}}}{\Pi_{i=1}^n (E_{i1})^{\frac{2}{n^2}}+\Pi_{i=1}^n (E_{i2})^{\frac{2}{n^2}}}}\nonumber\\
		\end{eqnarray}
	\end{itemize}

		For above measurement settings, classical post processing of $\overline{a_1}$ and choice of $s_1,s_2,...,s_n$ are such that the correlators are given by the following types of terms \cite{Tava}:
		\begin{eqnarray}
			C_{k_1,k_2}	&=&(\cos^{k_1}(t)\cdot \sin^{k_2}(t))\textmd{Tr}[(\sigma_1^{\otimes k_1}\otimes (\sigma_2^{\otimes k_2}))_{A_1}\otimes\nonumber\\
			&& (\sigma_1^{\otimes k_1}\otimes (\sigma_2^{\otimes k_2}))_{A_2,...,A_{n+1}}\cdot \otimes_{j=1}^n\rho_j]\,\,\textmd{with}\nonumber\\
			k_1+k_2&=&n,\,\, k_2=0\textmd{\small{ or even integer}}
		\end{eqnarray}
		For any fixed possible value of $k_1,k_2,$ variation of $\sigma_1,\sigma_2$ in the part of extreme parties' operators($ (\sigma_1^{\otimes k_1}\otimes (\sigma_2^{\otimes k_2}))_{A_2,...,A_{n+1}}$) gives different correlator terms.\\
		For example, let $n$$=$$3.$ Then there are $2^2$$=$$4$ correlators. Set $\{J_1,J_2,J_3,J_4\}$ is explicitly given by $\{C_{3,0},C_{3,2}\}.$ There are three different $C_{3,2}$ possible, say $C_{3,2}^{(1)},C_{3,2}^{(2)},C_{3,2}^{(3)}$ . Those are 
		\begin{itemize}
			\item $C_{3,2}^{(1)}$$=$$(\cos(t)\sin^{2}(t))\textmd{Tr}[(\sigma_1\otimes \sigma_2^{\otimes 2})_{A_1}\otimes
			(\sigma_1)_{A_2}\otimes (\sigma_2)_{A_3}\otimes (\sigma_2)_{A_4}\cdot \otimes_{j=1}^3\rho_j]$
			\item $C_{3,2}^{(2)}$$=$$(\cos(t)\sin^{2}(t))\textmd{Tr}[(\sigma_1\otimes \sigma_2^{\otimes 2})_{A_1}\otimes
			(\sigma_2)_{A_2}\otimes (\sigma_1)_{A_3}\otimes (\sigma_2)_{A_4}\cdot \otimes_{j=1}^3\rho_j]$
			\item $C_{3,2}^{(3)}$$=$$(\cos(t)\sin^{2}(t))\textmd{Tr}[(\sigma_1\otimes \sigma_2^{\otimes 2})_{A_1}\otimes
			(\sigma_2)_{A_2}\otimes (\sigma_2)_{A_3}\otimes (\sigma_1)_{A_4}\cdot \otimes_{j=1}^3\rho_j]$
		\end{itemize}

	Let $V_{\small{n-star}}$ denote the L.H.S of Eq.(\ref{ineqsn}). For above measurement settings, $V_{n-star}$ takes the form:
	\begin{eqnarray}\label{maxviost}
		V_{\small{n-star}}&=&\frac{\sum_{\substack{k_2\in\{0,1,2,...,n\}\\k_2\in\{0,\textmd{\small{even integer}}\leq n\}}}(\Pi_{i=1}^n (E_{i1})^{k_1}(E_{i2})^{k_2})^{\frac{1}{n^3}}G_{k_2}}{2^{n-2}\sqrt{\Pi_{i=1}^n (E_{i1})^{\frac{2}{n^2}}+\Pi_{i=1}^n (E_{i2})^{\frac{2}{n^2}}}}\nonumber\\
		&&\\
		G_{k_2}&=&\sum_{\substack{h_1,...,h_n\in\{1,2\}\\s.t. \Pi_{j=1}^{n}h_j=2^{k_2}}}\Pi_{i=1}^n(E_{ih_i})^{\frac{1}{n}}\,\textmd{\small{and} }k_1+k_2=n
	\end{eqnarray}	
	If $V_{n-star}$$>$$1,$ then Eq.(\ref{ineqsn}) is violated. Maximum value of $V_{n-star}$ is $\sqrt{2}.$ 
	
	To prove Theorem.\ref{theo3}, we will use $V_{n-star}$ and follow strategy similar to that used for proving Theorem.\ref{thinc1}. We first provide the condition over the set of $n$ two-qubit states used in the network for which Theorem.\ref{theo3} holds.
	\subsection*{Conditions over $\rho_1,\rho_2,...,\rho_n$} For our scenario, we consider that each of $n$ sources $\mathcal{S}_i$ is generating two-qubit entangled state $\rho_i$. Further, let $\rho_1,\rho_2,...,\rho_n$ satisfy the following criterion:
	\begin{eqnarray}\label{theoe04}
		H&<&(4\sum_{\substack{k_2\in\{0,1,2,...,n\}\\k_2:even}}(\Pi_{i=1}^n (E_{i1})^{k_1}(E_{i2})^{k_2})^{\frac{1}{n^3}}G_{k_2})^n\\
		&&\textmd{where}\nonumber\\
		H&=&2^{n^2}(\sqrt{\Pi_{i=1}^n (E_{i1})^{\frac{2}{n^2}}+\Pi_{i=1}^n (E_{i2})^{\frac{2}{n^2}}})\cdot\nonumber\\
		&&(\Pi_{i=1}^n (E_{i1})^{\frac{1}{n^2}}+\Pi_{i=1}^n (E_{i2})^{\frac{1}{n^2}})^{n-1}\,\,\textmd{\small{and}}\nonumber\\
		G_{k_2}&=&\sum_{\substack{h_1,...,h_n\in\{1,2\}\\s.t. \Pi_{j=1}^{n}h_j=2^{k_2}}}\Pi_{i=1}^n(E_{ih_i})^{\frac{1}{n}}
	\end{eqnarray}
	
	We next provide the incompatible measurement set for one of $n$ extreme parties.
	\subsection*{Fixed Incompatible Set of Measurements for One Extreme Party} 
	W.L.O.G., let $A_2$ perform fixed incompatible measurement settings. Let$A_2$ choose from following set of measurements:
	\begin{eqnarray}\label{incss5}
		x_{2,j}&=&\{\eta_1(\cos(t) \sigma_1+(-1)^j\sin(t) \sigma_2)\}_{j=0}^1,\nonumber
		\\
		\textmd{\small{where}}\,t&=&\arcsin\sqrt{\frac{\Pi_{i=1}^n (E_{i2})^{\frac{2}{n^2}}}{\Pi_{i=1}^n (E_{i1})^{\frac{2}{n^2}}+\Pi_{i=1}^n (E_{i2})^{\frac{2}{n^2}}}}\nonumber\\
		1&\geq&\eta_1>\frac{(\cos(t)+\sin(t))^{n-1}}{V_{\small{n-star}}^n}
	\end{eqnarray}
	This is the noisy version of measurement settings given by Eq.(\ref{binc1}). 
	Using Eq.(\ref{theoe04}), expression of argument $t$ from Eq.(\ref{incss5}) and that of $V_{\small{n-star}}$ from Eq.(\ref{maxviost}), we get:
	\begin{eqnarray}
		\frac{(\cos(t)+\sin(t))^{n-1}}{V_{\small{n-star}}^n}&<&1.
	\end{eqnarray}
	So range of $\eta_1$ given by Eq.(\ref{incss5}) is valid. Also, it is easy to see that for this given range $\eta_1$ automatically satisfies
	\begin{eqnarray}\label{etar}
		\eta_1&>&\frac{1}{\cos(t)+\sin(t)}\nonumber\\
	\end{eqnarray}
	By Eq.(\ref{etar}), it is clear that Eq.(\ref{incss5}) represents incompatible set of measurements for given range of $\eta_1.$\\
	Now that we have fixed the states used in the network along with the incompatible measurement settings of one of the extreme parties, we next complete the proof of Theorem.\ref{theo3} by searching for a set of compatible measurements for each of remaining $n$$-$$1$ extreme parties. 
	\subsection*{Finding Compatible Measurements for $A_3,...,A_{n+1}$} 
	We now complete proof of Theorem.\ref{theo3}. As per our requirement, we need to find out compatible measurements for $A_3,...,A_{n+1}$ such that $V_{\small{n-star}}$$>$$1.$ For that we use the same strategy as that used in proof of Theorem.\ref{thinc1}. \\
	$A_1$ performs GHZ basis measurement and $A_2$ performs from the incompatible measurement set given by Eq.(\ref{binc1}). We now need to find out a set of two compatible measurements for each of the extreme parties $A_3,..,A_{n+1}$ such that corresponding $n+1$-partite correlations violate $n$-local inequality(Eq(\ref{ineqsn})).\\
Now, as already pointed out before  that for given condition on the states(Eq.\ref{theoe04}), violation of Eq.(\ref{ineqsn}) occurs and hence non $n$-locality is detected in the network. So, given this class of states, we only need to search for compatible measurements for $A_3,..,A_{n+1}.$  \\
	$\forall i$$=$$3,4,...,n$$+$$1,$ let $A_{i}$ choose from the set $\{x_{i,0},x_{i,1}\}$ such that:
	\begin{eqnarray}\label{bsinc5}
		x_{i,j}&=&\{\eta_{i-1}(\cos (t) \sigma_1+(-1)^j\cos (t) \sigma_2)\}_{j=0}^1,\,\eta_{i-1}\in [0,1]\nonumber\\
		t&=&\arcsin\sqrt{\frac{\Pi_{i=1}^n (E_{i2})^{\frac{2}{n^2}}}{\Pi_{i=1}^n (E_{i1})^{\frac{2}{n^2}}+\Pi_{i=1}^n (E_{i2})^{\frac{2}{n^2}}}}
	\end{eqnarray}
	For these measurement settings, $V_{\small{n-star}}$*Eq.(\ref{maxviost}) gets modified to  $V_{\small{n-star}}^{(\eta_1,...\eta_n)}$(say). \\
	$V_{\small{n-star}}^{(\eta_1,...\eta_n)}$ is given by \cite{Bis}:
	\begin{eqnarray}\label{bsinc6}
		V_{\small{n-star}}^{(\eta_1,...\eta_n)}&=&\sqrt[n]{\Pi_{i=1}^n\eta_i}\cdot V_{\small{n-star}}
	\end{eqnarray}
	Set of measurement settings provided by Eq.(\ref{bsinc5}) is compatible if:
	\begin{equation}\label{binc3i}
		\eta_{i}\leq \frac{1}{\cos(t)+\sin(t)}\,\,\forall i=2,3,...,n
	\end{equation}
	As non $n$-locality correlations are detected in the network, Eq.(\ref{bsinc6}) implies that $\eta_1,..,\eta_n$ must satisfy:
	\begin{equation}\label{abinc5}
		\Pi_{i=2}^n	\eta_i>\frac{1}{\eta_1\cdot V_{\small{n-star}}^n}
	\end{equation} 
	Any collection of $\eta_2,...,\eta_n$ satisfying both Eq.(\ref{binc3i}) and Eq.(\ref{abinc5}) will suffice for our purpose. \\
	For each of $A_3,...,A_{n},$ let us set:
	\begin{equation}
		\eta_2=\eta_3=...=\eta_{n-1}=\frac{1}{\cos(t)+\sin(t)}.
	\end{equation}
	So each of $A_3,A_4,...,A_{n}$ perform compatible measurements. \\
	For above choice of measurements, we get from Eq.(\ref{abinc5}):
	\begin{equation}\label{ainc5}
		\eta_n>\frac{(\cos(t)+\sin(t))^{n-2}}{\eta_1\cdot V_{\small{n-star}}^n}
	\end{equation} 
	Now, by given conditions over $\eta_1$(Eq.\ref{incss5}), we have:
	\begin{eqnarray}
		\eta_1&>&\frac{(\cos(t)+\sin(t))^{n-1}} {V_{\small{n-star}}^n}\nonumber\\
		\Rightarrow	\frac{(\cos(t)+\sin(t))^{n-2}}{\eta_1\cdot V_{\small{n-star}}^n}&<&\frac{1}{(\cos(t)+\sin(t))}
	\end{eqnarray}
	Also,  $	\frac{(\cos(t)+\sin(t))^{n-2}}{\eta_1\cdot V_{\small{n-star}}^n}$$>$$0.$ 
	Hence, we get
	\begin{eqnarray}
		(\frac{(\cos(t)+\sin(t))^{n-2}}{\eta_1\cdot V_{\small{n-star}}^n},\frac{1}{\cos(t)+\sin(t)}]\nonumber\\
		\subset	[0,\frac{1}{\cos(t)+\sin(t)}].
	\end{eqnarray}
	
	Let $t^{'}$$\in$$(\frac{(\cos(t)+\sin(t))^{n-2}}{\eta_1\cdot V_{\small{n-star}}^n},\frac{1}{\cos(t)+\sin(t)}].$\\
	Setting $\eta_n$$=$$t^{'}$ in Eq.(\ref{bsinc5}) will suffice for our purpose.\\
	This completes our search of compatible measurement set for $A_{n+1}.$\\
	Theorem is thus proved$\blacksquare$
	\section{Appendix.F}\label{appnf}
	Here we will discuss the details of the numerical example(see main text) showing star topology giving advantage over linear topology in $4$-local network.\\
	Let us first consider linear $4$-local network.\\
	Let each of $\rho_1,\rho_2,\rho_3,\rho_4$ be a  Werner state with  visibility parameter $v_i$(say). So $E_{i,j}$$=$$v_i$, $\forall j$$=$$1,2,3.$ \\
	
	For these states, $B_{\small{4-lin}}$(Eq.(\ref{blin})) is given by:
	\begin{equation}\label{useles1}
		B_{\small{4-lin}}=\sqrt{2v_1v_2v_3v_4}=\sqrt{2V}
	\end{equation}
	Optimal measurement settings(for both extreme parties $A_1,A_5$) to achieve this bound is given by Eq.(\ref{ainc1}) for $r$$=$$\frac{\pi}{4}.$ These measurement settings are incompatible.\\
	Now let us consider our measurement contexts for the extreme parties:
	\begin{itemize}
		\item $A_1$ performs incompatible measurement 
		\item  $A_5$ performs compatible measurement 
	\end{itemize}
	Specifically, let $A_1,A_5$ perform:
	\begin{eqnarray}\label{fal1}
		x_{1,k}&=&\eta_1 \vec{n}_{1,k}\cdot \vec{\sigma}(k=0,1)\nonumber\\
			x_{5,k}&=&\eta_2 \vec{n}_{5,k}\cdot \vec{\sigma}(k=0,1)\nonumber\\
	\end{eqnarray}
	For these measurement settings of the extreme parties, let $I_4^{(\eta_1,\eta_2)},J_4^{(\eta_1,\eta_2)}$ denote the correlator terms appearing in $4$-local inequality. It is easy to check that these correlator terms are the scaled version of the correlator terms $I_4$ and $J_4$ with scaling factor $\eta_1\cdot\eta_2:$
	\begin{eqnarray}
		I_4^{(\eta_1,\eta_2)}&=&\eta_1\cdot \eta_2 I_4\nonumber\\
		J_4^{(\eta_1,\eta_2)}&=&\eta_1\cdot \eta_2 J_4\nonumber\\
	\end{eqnarray}
	The upper bound $B_{4-lin}^{(\eta_1,\eta_2)}$ thus takes the form:
	\begin{equation}\label{useles2}
		B_{\small{4-lin}}^{(\eta_1,\eta_2)}=\sqrt{2V\eta_1\cdot\eta_2}
	\end{equation}
	From the information about the optimal projective measurements for achieving the upper bound $	B_{\small{4-lin}}$(Eq.(\ref{useles1})), it is clear that to obtain the bound $B_{4-lin}^{(\eta_1,\eta_2)}$(Eq.(\ref{useles1})), the optimal projective measurements directions will be given $\vec{n}_{1,k}$$=$$\vec{n}_{5,k}$$=$$((-1)^k\frac{1}{\sqrt{2}},0,\frac{1}{\sqrt{2}})$in Eq.(\ref{fal1}) i,e, same as that in noiseless case with the noise parameters satisfying our measurement context of incompatible and compatible settings for $A_1$ and $A_5$ respectively:
	\begin{itemize}
		\item $\eta_1$$=$$1$ \\
		So, $x_{1,k}$$=$$\{\frac{1}{\sqrt{2}}(\sigma_3+(-1)^k \sigma_1)\}.$ Note that this a set of  two incompatible measurements(\ref{appna})
		\item $\eta_2$$=$$\frac{1}{\sqrt{2}}$ \\
		So, $x_{5,k}$$=$$\{\frac{1}{2}(\sigma_3+(-1)^k \sigma_1)\}.$ Note that this a set of  two compatible measurements(\ref{appna}).
	\end{itemize}
	For the above optimal projective measurement settings the bound $	B_{\small{4-lin}}^{(1,\frac{1}{\sqrt2})}$(Eq.(\ref{useles2})) of $4$-local inequality
	\begin{eqnarray}\label{useles3}
		B_{\small{4-lin}}^{(1,\frac{1}{\sqrt2})}&&=\sqrt{2 V\frac{1}{\sqrt{2}}}\nonumber\\
		&=&\sqrt{\sqrt{2} V}
	\end{eqnarray}
	Now for our numerical example, we have considered identical Werner states: $v_i$$=$$0.74,$ $\forall i$$=$$1,2,3,4.$\\
	From Eq.(\ref{useles3}), we get:
	$B_{\small{4-lin}}^{(1,\frac{1}{\sqrt2})}$$=$$0.6512.$
	So $4$-local inequality is not violated for optimal projective measurement context where $A_1$ is performing from a set of two incompatible dichotomic measurements and $A_5$ performing from a set of two compatible dichotomic measurements.

	Let us now consider $4$-local star network. We use same $\rho_1,\rho_2,\rho_3,\rho_4$ as considered in linear network.\\
	Let $A_2$ perform following incompatible measurements:
	\begin{equation}
		x_{2,j}=\{\frac{1}{\sqrt{2}}(\sigma_1+(-1)^j \sigma_2)\}\,\,j=0,1
	\end{equation}
	Let each of remaining three extreme parties perform following compatible measurements:
	\begin{equation}
		x_{3,j}=	x_{4,j}=x_{5,j}=\{0.5(\sigma_1+(-1)^j\sigma_2 )\}\,\,j=0,1
	\end{equation}
	These settings belong to the class of measurement settings for which the bound of $n$-local inequality(Eq.(\ref{ineqsn})) is derived(\ref{appne}). 
	For these settings, $V_{4-star}$(Eq.(\ref{maxviost})) gives value $1.0114.$ So  Eq.(\ref{ineqsn}) is violated.\\
	Consequently non $4$-local correlations are detected in $4$-local star network but not in linear $4$-local network inspite of using optimal measurements in linear topology.
\section{Appendix.G}\label{appng}
\textit{Proof of Theorem.\ref{theo30}:} Each of $n$ independent sources $\mathcal{S}_1,...,\mathcal{S}_n$ is distributing arbitrary two-qubit entangled state $\rho_1,...,\rho_n$ respectively.
Let us consider the following measurement settings:
\begin{itemize}
	\item $	x_{i,0}$$=$$\sigma_1$; $x_{i,1}$$=$$\sigma_2$ $\forall i$$=$$2,3,...,n$$+$$1.$
	\item $A_1$ performs GHZ basis measurement.
\end{itemize}
Let $\mathcal{C}_{star}$$=$$\{J_1,J_2,...,J_{2^{n-1}}\}$ denote the collection of all correlator terms appearing in Eq.(\ref{ineqsn}).
For above measurement settings, $\mathcal{C}_{star}$ is given as follows:
\begin{eqnarray}\label{appf1}
	\mathcal{C}_{star}&=&\frac{1}{2^n}\{\Pi_{1}^{n}(E_{ih_i})\}_{\substack{h_1,...,h_n\in\{1,2\}\\s.t. \Pi_{j=1}^{n}h_j=2^{k}\\}},\\
	&&\forall k\in\{0,\textmd{even integer}\leq n\}
\end{eqnarray}
Simplifying  above form of correlators(Eq.(\ref{appf1})), L.H.S. of Eq.(\ref{ineqsn}) turns out to be:
\begin{eqnarray*}
	\sum_{i=1}^n	\sqrt[n]{|J_i|}&=&\sum_{\substack{k\in\{0,1,2,...,n\}\\k\in\{0,\textmd{\small{even integer}}\leq n\}}}\sum_{\substack{h_1,...,h_n\in\{1,2\}\\s.t. \Pi_{j=1}^{n}h_j=2^{k}}}\frac{\Pi_{i=1}^n(E_{ih_i})^{\frac{1}{n}}}{2^{n-1}}\\
	&\leq& 1
\end{eqnarray*}
Hence, $n$-local inequality(Eq.(\ref{ineqsn})) is not violated for any $\rho_1,\rho_2,...,\rho_n.$
\section{Appendix.H}\label{appnh}
Before we prove Theorem.\ref{corr1}, we first discuss about full network  non $n$-local correlations \cite{Poz}.
\subsection*{Full Network Non $n$-local Correlations}
As introduced in \cite{Poz}, for any given measurement scenario, network correlations are said to be \textit{fully network nonlocal} if and only if we cannot model the correlations in terms of a $n$-local hidden variable(HV) model such that at least one source in the network is of a local-variable nature whereas all the remaining sources, in general, can be independent nonlocal resources.\\
Let us consider a $n$-local network. W.L.O.G., let us fix the measurement scenario corresponding to $n$-local star network as the measurement scenario here(one may however consider any other measurement scenario).\\
Precisely central party $A_1$ performing single measurement whereas each of $n$ edge parties($A_2,..,A_{n+1}$) performing from a set of two arbitrary dichotomic projective measurements. \\
Corresponding measurement correlation $p(\bar{a}_1,a_2,...,a_{n+1}|x_2,...,x_{n+1})$ is \textit{not full network nonlocal} if it can be decomposed as:
\begin{eqnarray}\label{fnn1}
	P(\bar{a}_1,a_2,...,a_{n+1}|x_2,...,x_{n+1})=\sum_{\lambda}\sigma(\lambda)
	P(a_j|x_j,\lambda)Q\,\,\textmd{where}\nonumber\\
	Q=P(\bar{a}_1,a_2,...a_{j-1},a_{j+1},...,a_{n+1}|x_2,..,x_{j-1},x_{j+1},...,x_{n+1},\lambda)\nonumber	\\
\end{eqnarray}
$\sigma(\lambda)$ denote probability distribution of the local hidden variable $\lambda$ characterizing $j^{th}$ source $\textbf{S}_j$ shared between $A_1$ and $A_j.$\\
Eq.(\ref{fnn1}) points out that for any $j$$\in$$\{1,2,,...,n$\},$ j^{th}$ source is characterized by a local hidden variable $\lambda.$ So it is clear that if at least one of $n$ sources can be modeled by a local hidden variable, then even if all remaining $n$$-$$1$ sources are maximally nonlocal(can be modeled by bipartite no-signalling box), corresponding network correlations fail to be fully network nonlocal.\\
We now prove the theorem\ref{corr1}.
\subsection*{Proof of Theorem.\ref{corr1}}
As already said above, we are considering measurement context corresponding to star $n$-local network for our purpose. For any other measurement scenario, the theorem can be proved similarly.\\
W.L.O.G. let edge party $A_2$ perform from a set of two compatible measurements whereas remaining $n$$-$$1$ edge parties perform from a set of incompatible measurements and central party perform single measurement. Let $\{x_{k,0},x_{k,1}\}$ denote the set of measurements for $k^{th}$ edge party $\forall k$$=$$2,3,,...,n$$+$$1$.\\
If we can show that resulting measurement correlations can be written in form given by Eq.(\ref{fnn1})(for $j$$=$$2$), then that completes the proof.\\
For ease of writing, using $(a_i,x_i)$ for labeling any (output,input) pair of $A_i$($i$$=$$2,3,,...,n$$+$$1)$. 
$\forall i$$=$$2,3,,...,n$$+$$1,$ let $M_i^{(a_i|x_i)}$ denote POVM element corresponding to the (output,input) pair $(a_i,x_i).$\\
As $A_2$ perform compatible measurements, so there exists a parent POVM $\{G_{2,\lambda}\}_{\lambda}$ such that:
\begin{eqnarray}\label{fnn2}
	M_{2}^{a_2|x_{2}}&=&\sum_{\lambda} P(a_2|x_{2},\lambda) G_{2,\lambda},\,\forall x_2,a_2\\
\end{eqnarray}
Let $\bar{a}_1$ denote two bit output string of $A_1:$ $\bar{a}_1$$=$$(a_{11},a_{12},....,a_{12^n})$ with $a_{21},a_{22}$$\in$$\{0,1\}.$ Let $M_{\bar{a}_1}$ denote POVM element corresponding to any output bit string $\bar{a}_1$ of $A_1$.\\
Let $\mathcal{S}_i$ distribute arbitrary two-qubit state $\rho_i$ $\forall i$$=$$1,2,,...,n$. \\
With these notations, we can write any probability term $P(\bar{a}_1,a_2,...,a_{n+1}|x_2,...,x_{n+1})$ as:
\begin{eqnarray}\label{fnn3}
	P(\bar{a}_1,a_2,...,a_{n+1}|x_2,...,x_{n+1})=\textmd{Tr}[( M_{\bar{a}_1}\otimes M_{2}^{a_2|x_2}\otimes...\nonumber\\
	\otimes M_{n+1}^{a_{n+1}|x_{n+1}} ) \cdot \otimes_{i=1}^n\rho_i]\nonumber\\
	=\sum_{\lambda}  P(a_2|x_{2},\lambda) \textmd{Tr}[( M_{\bar{a}_1}\otimes G_{2,\lambda}\otimes  M_{3}^{a_{3}|x_{3}} \otimes...\nonumber\\
	\otimes M_{n+1}^{a_{n+1}|x_{n+1}} ) \cdot \otimes_{i=1}^n]\nonumber\\
	=\sum_{\lambda}  P(a_2|x_{2},\lambda)\cdot P(\bar{a}_1,a_3,...,a_{n+1}|x_3,...,x_{n+1},\lambda)\nonumber\\
	=\sum_{\lambda}   P(\lambda)\cdot P(a_2|x_{2},\lambda)\cdot P(\bar{a}_1,a_3,...,a_{n+1}|x_3,...,x_{n+1},\lambda)\nonumber\\
	\\
\end{eqnarray}
This proves the theorem.


\begin{thebibliography}{1}	
		
		\bibitem{Bci} C. Branciard, N. Gisin, and S. Pironio, "\textbf{Characterizing the Nonlocal Correlations Created via Entanglement Swapping}", Phys. Rev. Lett. \textbf{104}, 170401 (2010).
		
		\bibitem{Bran} C. Branciard, D. Rosset, N. Gisin
		and S. Pironio,	"\textbf{Bilocal versus non-bilocal correlations in entanglement swapping experiments}", Phys. Rev. A 85, 032119 (2012). 
		
		\bibitem{Fri} T. Fritz, "\textbf{Beyond Bell's theorem: correlation scenarios}", New J. Phys. \textbf{14} 103001 (2012). 
		
		
		\bibitem{Ros} D. Rosset,\emph{etal.}, "\textbf{Nonlinear bell inequalities tailored for quantum networks}", Phys. Rev. Lett. \textbf{116}, 010403 (2016).		
			
		
		\bibitem{Cha} R. Chaves, "\textbf{Polynomial bell inequalities}", Phys. Rev. Lett.
		\textbf{116}, 010402 (2016).
		
		\bibitem{Ren}, M.-O. Renou, E. Baumer, S. Boreiri, N. Brunner, N. Gisin, and S. Beigi, "\textbf{Genuine Quantum Nonlocality in the Triangle Network}"  Phys. Rev. Lett. \textbf{123}, 140401 (2019).
		
		
	\bibitem{Reno} M.-O. Renou,\emph{etal.}, "\textbf{Quantum theory based on real numbers can be experimentally falsified}" Nature \textbf{600}, 625 (2021).
		
		\bibitem{Kauu} K. Mukherjee, B. Paul, and D. Sarkar, "\textbf{Correlations In n-local Scenario}", Quantum Inf Process 14, 2025–2042 (2015).
		
		
		\bibitem{ujjwal} P. Ghosh, C. Srivastava, S. Choudhary, U. Sen, "\textbf{Measurement incompatibility at all remote parties do not always permit Bell nonlocality}", Phys. Rev. A \textbf{111}, 052208 (2025).
		
		\bibitem{Kas}  K. Mukherjee, B. Paul, and D. Sarkar, "\textbf{Nontrilocality: Exploiting nonlocality from three particle systems}", Phys. Rev. A
		\textbf{96}, 022103 (2017).
		
		\bibitem{Muk} K. Mukherjee, B. Paul, and D. Sarkar, "\textbf{Characterizing quantum correlations in a fixed-input n-local network scenario}", Phys. Rev. A \textbf{101}, 032328 (2020).
		
		\bibitem{Taav} A. Tavakoli, A. Pozas-Kerstjens, M-X Luo and M-O Renou	"\textbf{Bell nonlocality in networks}", Reports on Progress in Physics, Volume \textbf{85}, Number 5 (2022).
		
		\bibitem{Cla} J. F. Clauser, M. A. Horne, A. Shimony, and R. A. Holt, "\textbf{Proposed Experiment to Test Local Hidden-Variable Theories}", Phys. Rev. Lett. \textbf{23}, 880 (1969).
		
		\bibitem{Gis} N. Gisin, Q.X. Mei, A. Tavakoli,M.O. Renou, and N.  Brunner, "\textbf{All entangled pure quantum states violate the bilocality inequality}" Phys. Rev. A \textbf{96}(2), 020304 (2017).
		
		
		\bibitem{And} F. Andreoli, G. Carvacho, L. Santodonato, R. Chaves and F. Sciarrino", "\textbf{Maximal qubit violation of n-locality inequalities in a star-shaped quantum network}", New J. Phys. 19 113020 (2017).
		
		\bibitem{Tavak} A. Tavakoli, N. Gisin, and Cyril Branciard, "\textbf{Bilocal Bell inequalities violated by the quantum Elegant Joint Measurement}" Phys. Rev. Lett. \textbf{126}, 220401 (2021).
		
		
		
		\bibitem{Poz} A. Pozas-Kerstjens, N. Gisin and A. Tavakoli, "\textbf{Full Network Nonlocality}", Phys. Rev. Lett. 128, 010403 (2022).
		
		
		
		\bibitem{Hein} T. Heinosaari, T. Miyadera, and M. Ziman, An Invitation to Quantum Incompatibility, Journal of Physics A: Mathematical and Theoretical \textbf{49}, 123001 (2015).
		
		
		\bibitem{Guh} O. Guhne, E. Haapasalo, T. Kraft, J-P Pellonpaa
		, and R. Uola, "\textbf{Incompatible measurements in quantum information science}", Rev. Mod. Phys. \textbf{95}, 011003(2023).
		
		\bibitem{Fin} A. Fine, "\textbf{Hidden Variables, Joint Probability, and the Bell Inequalities}", Phys. Rev. Lett. \textbf{48}, 291 (1982).
		
		\bibitem{Wol} M. M. Wolf, D. Perez-Garcia, and C. Fernandez, "\textbf{Measurements incompatible in quantum theory cannot be measured jointly in any other no-signaling theory}", Physical Review Letters \textbf{103}, 230402 (2009).
		
			\bibitem{Qui} M. T. Quintino, J. Bowles, F. Hirsch, and N. Brunner, "\textbf{Incompatible quantum measurements admitting a local hidden variable model}", Physical Review A \textbf{93}, 052115 (2016). 
			\bibitem{Hir} F. Hirsch, M. T. Quintino, and N. Brunner, "\textbf{Quantum measurement incompatibility does not imply Bell nonlocality}",Physical Review A \textbf{97}, 012129 (2018).
				
				\bibitem{Ben} E. Bene and T. Vértesi, "\textbf{Measurement incompatibility does not give rise to Bell violation in general}", New Journal of Physics \textbf{20}, 013021 (2018).
				
				\bibitem{Benn} C. H. Bennett, D. P. DiVincenzo, J. A. Smolin, and W. K. Wootters, "\textbf{Mixed-state entanglement and quantum error correction}", Phys. Rev. A \textbf{54}, 3824 (1996).
		
		\bibitem{Hei} T. Heinosaari, J. Kiukas and D. Reitzner, \textbf{Noise robustness of the incompatibility of quantum measurements}, Phys. Rev. A \textbf{92}, 022115 (2015).
		
		
		\bibitem{Gue} L. Guerini, J. Bavaresco, M. Terra Cunha, and A. Acin, \textbf{Operational framework for quantum measurement simulability}, J. Math. Phys. \textbf{58}, 092102 (2017) 
		
		\bibitem{Sky} P. Skrzypczyk1, I. Supic and D. Cavalcanti, \textbf{All Sets of Incompatible Measurements give an Advantage in Quantum State Discrimination}, Phys. Rev. Lett. \textbf{122}, 130403 (2019).
		
			\bibitem{Tava} A. Tavakoli, P. Skrzypczyk, D. Cavalcanti, and A. Acín, "\textbf{Nonlocal correlations in the star-network configuration}", Phys. Rev. A \textbf{90(6)}, 062109 (2014)
		
		\bibitem{rece} K.Mukherjee and B.Paul, "\textbf{Network Nonlocality Without Entanglement Of All Sources}", arxiv. 2410.15131 (2024).
		
			
		\bibitem{Bus}P. Busch, "\textbf{Unsharp reality and joint measurements for spin observables}",  Phys. Rev. D \textbf{33}, 2253(1986).
		
		\bibitem{Bis} K.Mukherjee, I.Chakrabarty and G.Mylavarapu, "\textbf{ Persistency of non-n-local correlations in noisy linear networks}", Phys. Rev. A \textbf{107},032404 (2023). 
		
		\bibitem{Luo} S. Luo, "\textbf{Quantum discord for two-qubit systems}", Phys. Rev. A \textbf{77}, 042303 (2008).
		
		
		\bibitem{Bru} N. Brunner, D. Cavalcanti, S. Pironio, V. Scarani, and S. Wehner "\textbf{Bell nonlocality}", Rev. Mod. Phys. \textbf{86}, 419 (2014).
		
		\bibitem{recent}L. Tendick, C.Budroni and M.T. Quintino. "\textbf{Strict hierarchy between n-wise measurement simulability, compatibility structures, and multi-copy
			compatibility}", arxiv:2506.21223v2 (2025).
		
	\end{thebibliography}
\end{document}